\documentstyle[preprint,aps]{revtex}

\begin{document}

\draft

\preprint{January,2001}

\title{Toy model for  a two-dimensional accretion disk
\\dominated by Poynting flux}

\author{Hyun Kyu Lee\footnote{e-mail : hklee@hepth.hanyang.ac.kr}}

\address{Department of Physics, Hanyang University, Seoul, 133-791, Korea \\
and \\ Asia Pacific Center for Theoretical Physics, Seoul 130-012,
Korea}

\maketitle

\begin{abstract}
We discuss the effect of the Poynting flow  on the magnetically
dominated thin  accretion disk, which is simplified to a
two-dimensional disk on the equatorial plane.  It is shown in the
relativistic formulation that the Poynting flux by the rotating
magnetic field with Keplerian angular velocity can balance the
energy and angular momentum conservation of a steady accretion
flow.
\end{abstract}


\narrowtext

\section{Introduction}

The Poynting flux model\cite{lovelace}\cite{blandford} suggested
for an accretion disk with the ordered magnetic fields has been
considered to be one of the viable models for the astrophysical
jets\cite{bbr}\cite{love}.  In contrast to the hydrodynamic jets
in which the energy and angular momentum are carried by the
kinetic flux of the matter, the Poynting flux is characterized by
the outflow of energy and angular momentum carried predominantly
by the electromagnetic field.

In the non-relativistic formulation, Blandford\cite{blandford}
suggested an
 axisymmetric and steady solution  for the Poynting outflow
assuming a force-free magnetosphere surrounding an accretion disk.
The poloidal field configuration for a black hole in a force-free
magnetosphere has been discussed recently by Ghosh\cite{ghosh} in
the relaticistic formulation, in which the possible forms of the
poloidal magnetic fields are suggested.
 The developments  of the
ordered magnetic field and the Poynting outflow on the disk have
been studied by many authors\cite{ga}\cite{lop}\cite{fendt} and
recently Ustyugova et al.\cite{love} perform a  axisymmetric
magnetohydrodynamical simulation to show that the quasi-stationary
and approximately force-free Poynting jet from the inner part of
the accretion disk is possible.

The Poynting flux in a system of black hole-accretion disk
recently has also been studied in connection to the gamma ray
bursts\cite{lwb}\cite{lbw}\cite{li1}. One of the main reasons is
that the Poynting flux carries very small baryonic component,
which is essential for powering the gamma ray bursts\cite{piran}.
The evolution of the system is found to be largely depend on the
Poynting flux on the disk\cite{lk}\cite{li}. The relativistic
effects on the accreting flows close to a black hole has been
discussed by Lasota\cite{lasota}, Abramowicz et al.\cite{acgl} and
recently by Gammie and Popham\cite{gp} in which the relativistic
effects on the slim-disk with the vertical structures averaged are
discussed in detail with viscous stresses. However the effect of
relativity on the accreting flow dominated by the Poynting flux
has not been discussed well in depth so far.

The purpose of this work is to study the effect of the Poynting
flux on the accretion flow in the relativistic formulation. In
this work, we consider a toy model for a magnetically dominated
thin accretion disk, which is assumed to be a two-dimensional disk
located on the equatorial plane with a black hole at the center.
To see the magnetic effect transparently it is also assumed that
there is no viscous stress tensor in the disk and there is no
radiative transfer from the disk.  We develop a relativistic
description of the two-dimensional model for a magnetically
dominated thin accretion disk in the background Kerr geometry. It
is shown that  the energy balance of the accretion disk for a
stationary accretion flow can be maintained by the Poynting flux,
provided that  the poloidal magnetic field is rotating with the
same Keplerian angular velocity $\Omega_K $  on the disk.

\section{Two-dimensional accretion flow}

The stress-energy tensor $T^{\mu\nu}_m$ of the matter in the disk
can be  given in the general form
\begin{eqnarray}
T^{\mu\nu}_{m} = (\rho_m + p + \Pi)u^{\mu}u^{\nu} + pg^{\mu\nu} +
S^{\mu\nu} +  u^{\mu}q^{\nu} + u^{\nu}q^{\mu}, \end{eqnarray}
where $\rho_m, \Pi,$ and $ p$ are  the rest-mass density, internal
energy and pressure respectively.  $S^{\mu\nu}$ is the viscous
tensor and $q^{\mu}$ is the radiative energy flux\cite{acgl}.

In this work we simplify the accretion disk to be
non-viscous($S^{\mu\nu}=0$), cool($p=0,  \Pi=0$) and
non-radiative($q^{\mu}=0$) disk to investigate the effect of the
magnetic field transparently. It is also assumed that there is
negligible mass flow in the direction perpendicular to the disk:
$u^{\theta} =0$. Then the stress-energy tensor is given simply by
\begin{eqnarray}
T^{\mu\nu}_{m} = \rho_m u^{\mu}u^{\nu}. \end{eqnarray} The energy
flux is obtained by
\begin{eqnarray}
{\cal E}^{\mu}_{(m)} = -\rho_m u_0 u^{\mu},
\label{Em}\end{eqnarray} and the angular momentum flux by
\begin{eqnarray}
{\cal L}^{\mu}_{(m)} = \rho_m u_{\phi} u^{\mu}.
\label{Lm}\end{eqnarray} For an idealized thin disk on
two-dimensional plane we take
\begin{eqnarray}
\rho_m = \frac{\sigma_m}{\rho} \delta(\theta-\pi/2),
\end{eqnarray} where $\sigma_m$ is the surface rest-mass density.
In this work, the background geometry is assumed to be  Kerr
metric(see Appendix A) with rotating black hole at the center and
we adopt the natural unit $G = c =1$.

The rate of the rest-mass flow crossing the circle  of radius $r$
is given by
\begin{eqnarray}
\int_r \alpha \rho_m u^r \frac{\rho^2 \varpi}{\sqrt{\Delta}}
d\theta d \phi =
 2\pi \sigma_m \rho u^r,
 \end{eqnarray}
which defines   the mass accretion rate $\dot{M}_+$  by
\begin{eqnarray}
\dot{M}_+ = - 2\pi \sigma_m \rho u^r \label{M+0}.\end{eqnarray}
This expression is identical to that derived in \cite{acgl} and
\cite{gp}, in which the vertical structures are averaged. The
stationary accretion flow implies that  $\dot{M}_+$ is
$r$-independent.

Using eq.(\ref{Em}) and (\ref{Lm}) the radial flow of the matter
energy at $r$ can be  given by
\begin{eqnarray}
\int_r \alpha {\cal E}^{r}_{(m)} \frac{\rho^2
\varpi}{\sqrt{\Delta}} d\theta d \phi= - 2\pi \sigma_m \rho u^r ~
u_0 = u_0 \dot{M}_+ \label{emr} \end{eqnarray} and the radial flow
of the matter angular momentum by
\begin{eqnarray}
\int_r \alpha {\cal L}^{r}_{(m)} \frac{\rho^2
\varpi}{\sqrt{\Delta}} d\theta d \phi
=
2\pi \sigma_m \rho u^r ~ u_{\phi} =- u_{\phi}~ \dot{M}_+
.\label{lmr} \end{eqnarray}

For the magnetosphere  outside the accretion disk, there are
electromagnetic  currents, $J^{\mu}$, which can flow through the
central object at the center and also along the magnetic field
lines anchored on the disk. In this work we  suppose  a situation
in which the currents flow into the inner edge of the disk through
the central object and the currents flows out along the magnetic
field lines from the disk as discussed by Blandford and Znajek
\cite{blandford}\cite{BZ}. The continuity of the currents(current
conservation) necessarily requires  surface currents in radial
direction on the disk. Similar consideration has been applied to
the black hole horizon as discussed by Thorne et al.\cite{TPM}.

In general the structure of the electromagnetic field with
discontinuity at $\theta= \pi/2$ plane can be reproduced by
assigning the surface charge and current on the plane in addition
to the bulk charge and current distributions which terminate on
the disk. The surface charge density $\sigma_e$ and surface
current density $K^{\hat{i}}$ can be defined systematically using
the procedure suggested by Damour\cite{damour} (See Appendix B for
details):
\begin{eqnarray}
\sigma_e = -\frac{E^{\hat{\theta}}}{4\pi},\label{eE} \\
K^{\hat{r}} = -\frac{1}{4\pi} B^{\hat{\phi}}, ~~~ K^{\hat{\phi}}
=\frac{1}{4 \pi} B^{\hat{r}}.\label{kbd} \end{eqnarray}

\section{Poynting flux}

The energy and angular momentum of the disk can be carried out by
Poynting flux  along the magnetic field lines anchored on the
disk. In our simple model, Poynting flux is the main driving force
for the accretion flow. We calculate the Poynting flux on a
two-dimensional disk  assuming a force-free magnetosphere around
the accretion disk.

Using the Killing vector in $t$-direction,
\begin{eqnarray}
\xi^{\mu} = (1,\, 0,\, 0,\,0),
 \end{eqnarray}
  we can define the energy flux
${\cal E}^{\mu}$ from the energy momentum tensor, $T^{\mu\nu}$
\begin{eqnarray}
{\cal E}^{\mu} &=& - T^{\mu\nu}\xi_{\nu} = (\alpha^2 -\varpi^2
\beta) T^{\mu 0} -\varpi^2\beta T^{\mu\phi},\\
{\cal E}^{\mu}_{~~ ;\mu} &=& 0, \end{eqnarray} where
\begin{eqnarray}
T^{\mu\nu} = \frac{1}{4\pi}(F^{\mu}_{~ \rho}F^{\nu\rho}
-\frac{1}{4}
g^{\mu\nu}F_{\rho\sigma}F^{\rho\sigma}).\label{tmunu1}
 \end{eqnarray}

Since we are interested in $\theta$-direction (normal to the disk
on the equatorial plane ), we consider ${\cal E}^{\theta}$ for
which the second term in eq.(\ref{tmunu1}) vanishes:
\begin{eqnarray}
{\cal E}^{\theta}=  \frac{1}{4\pi}(\alpha^2 -\varpi^2
\beta)F^{\theta}_{~ \rho}F^{0\rho}-
\frac{1}{4\pi}\varpi^2\beta F^{\theta}_{~ \rho}F^{\phi \rho}.
\end{eqnarray} Using the identities given in Appendix A, it can be
rewritten in terms of electric and magnetic fields as given  by
\begin{eqnarray}
{\cal E}^{\theta} &=& \frac{1}{4\pi \rho}[\alpha \hat{E} \times
\hat{B} |_{\theta} + \beta \varpi (E^{\hat{\theta}}E^{\hat{\phi}}+
B^{\hat{\phi}}B^{\hat{\theta}})].\label{erhatd} \end{eqnarray} The
power measured at infinity can be obtained using eq.(\ref{erhatd})
by
\begin{eqnarray}
P^{\theta}_{energy} &=& \int_{\theta =\pi/2} \alpha{\cal
E}^{\theta} d\Sigma_{\theta}
 = \int
\frac{\alpha}{4\pi}[\alpha (\vec{E} \times \vec{B})_{\theta} +
\varpi\beta(E^{\hat{\theta}}E^{\hat{\phi}}+
B^{\hat{\phi}}B^{\hat{\theta}})] \frac{\rho \varpi}{\sqrt{\Delta}}
d r d \phi. \label{powerz}\end{eqnarray} When the tangential
components of the electromagnetic field are multiplied by the laps
function $\alpha$ as in the membrane paradigm\cite{TPM}, we can
hide $\alpha$ in the integrand.

For the steady and axisymmetric case which we are interested in ,
$E^{\hat{\phi}} =0$, we get
\begin{eqnarray}
{\cal E}^{\theta} &=& \frac{1}{4\pi \rho}(-\alpha E^{\hat{r}}
B^{\hat{\phi}} + \beta \varpi B^{\hat{\phi}}B^{\hat{\theta}}).
\label{erhatd1} \end{eqnarray} The first term in the integrand in
eq.(\ref{powerz}) can be rewritten in terms of the surface current
density: Using eq.(\ref{kbad}) we get
\begin{eqnarray}
\frac{1}{4\pi}(\vec{E} \times \vec{B})_{\theta} = \vec{E}\cdot
\vec{K}.
 \end{eqnarray}
For the current in the direction of tangential electric field on
the disk  there is a energy dissipation into the disk surface.
This is what one can expect on the black hole horizon\cite{TPM}.
However for the current in the opposite direction this term
corresponds to the electro-motive force and it is the case for the
accretion disk on the equatorial plane discussed in Blandford and
Znajek\cite{blandford}\cite{BZ}as well as in this work.

Similarly the third term in eq.(\ref{powerz})(equivalently the
second term in eq.(\ref{erhatd1})) can be rewritten in terms of
the surface current using eq.(\ref{kbad}) as given by
\begin{eqnarray}
\frac{1}{4\pi} \varpi\beta B^{\hat{\phi}}B^{\hat{\theta}} = \varpi
\omega K^{\hat{r}} B^{\hat{\theta}}. \end{eqnarray}
 We can see  that this is the
magnetic braking power on the rotating body with an angular
velocity $\omega = - \beta$.

 Using the Killing vector in
$\phi$-direction for the axial symmetric case,
\begin{eqnarray}
\eta^{\mu} = (0,\, 0,\, 0,\,1),
 \end{eqnarray}
  we can define the angular momentum flux
${\cal L}^{\mu}$ from the energy momentum tensor, $T^{\mu\nu}$,
\begin{eqnarray}
{\cal L}^{\mu} &=& T^{\mu\nu}\eta_{\nu}= \varpi^2\beta T^{\mu 0} +
\varpi^2T^{\mu\phi},
\\ {\cal
L}^{\mu}_{~~ ;\mu} &=& 0,
 \end{eqnarray}

Then  we get the flux in $\theta$-direction (normal to the disk on
the equatorial plane ), ${\cal L}^{\theta}$,
\begin{eqnarray}
{\cal L}^{\theta} = -\frac{\varpi}{4\pi
\rho}B^{\hat{\theta}}B^{\hat{\phi}}\label{ltheta} =
\frac{1}{\rho}\varpi K^{\hat{r}} B^{\hat{\theta}},  \end{eqnarray}
which is nothing but a torque applied on the surface current
density $K^{r}$. Eq.(\ref{erhatd}) can be written as
\begin{eqnarray}
{\cal E}^{\theta} &=& \frac{1}{4\pi \rho}\alpha \hat{E} \times
\hat{B} |_{\theta} - \beta {\cal L}^{\theta}. \label{erhatdl}
\end{eqnarray}

In this work we assume that there is a sufficient ambient plasma
around the disk to maintain the force-free
magnetosphere\cite{blandford}\cite{love}. The force-free condition
for the magnetosphere with the current density $J^{\mu}$ is given
by
\begin{eqnarray}
F_{\mu\nu}J^{\mu} = 0, \label{ff}
 \end{eqnarray}
The magnetic flux, $\Psi$, through a circuit encircling
 $ \phi = 0 \rightarrow 2 \pi$,
\begin{eqnarray}
\Psi = \oint  A_{\phi} d\phi = 2\pi A_{\phi}, \label{psiaphi}
\end{eqnarray}
 defines a magnetic surface on which  $A_{\phi}(r,\theta) $ is
constant and therefore is characterized by the magnetic flux
$\Psi$ contained inside it.  From the  force-free condition it can
be shown that $A_0 $ is also constant along the magnetic field
lines and the electric field is always perpendicular to the
magnetic surface. We can define a function $\Omega_F(r,\theta)$,
\begin{eqnarray}
dA_0 = -\Omega_F dA_{\phi}, \label{omegaa}\end{eqnarray} which is
also constant along the magnetic surface. $\Omega_F$ can be
identified as an angular velocity
 of magnetic field line on a magnetic surface\cite{BZ}\cite{TPM}.

Then we get from the force-free condition
\begin{eqnarray}
\vec{E} = -\frac{\varpi}{\alpha}(\Omega_F +
\beta) \hat{\phi} \times \vec{B}, \label{epa}
\end{eqnarray} and  the Poynting flux perpendicular to the disk can be
written as
\begin{eqnarray}
{\cal E}^{\theta} &=& -\frac{\Omega_F \varpi}{4\pi
\rho}B^{\hat{\theta}}B^{\hat{\phi}} = \Omega_F {\cal
L}^{\theta}_{(\phi)}, \label{erhatd2} \end{eqnarray} as in
\cite{BZ}.

\section{Energy and angular momentum conservation}

In the idealized thin disk considered in this work, it is assumed
that  there is no radiative transfer or viscous interaction which
balance the radial flow of the  energy and angular momentum of the
accreting matter. Hence the electromagnetic field anchored on the
disk is responsible for the conservation of the total  energy and
angular momentum. Since we assume the idealized two-dimensional
disk, there is no radial flow of electromagnetic energy and
angular momentum unless there is any singular structure of
electromagnetic field on the disk.

Let us consider a circular strip at $r$ with infinitesimally small
width $\delta r$. Using eq.(\ref{ltheta}) the angular momentum
flux of electromagnetic field into the $\theta$-direction is given
by
\begin{eqnarray}
\Delta L = 2 \int^{r + \delta r}_r \alpha {\cal L}^{\theta}
d\Sigma_{\theta} = - B^{\hat{\theta}}B^{\hat{\phi}}\rho\varpi
\delta r , \end{eqnarray} where the factor 2 is introduced to take
account of both sides of the disk. It  is balanced by the change
of the angular momentum of matter given by
\begin{eqnarray}
\Delta L_m = - u_{\phi}\dot{M}_+|^{ r + \delta r}_{r} \rightarrow
-\frac{d u_{\phi}}{dr}\dot{M}_+ \delta r.
 \end{eqnarray}

Then we get
\begin{eqnarray}
\dot{M}_+ &=& \frac{B^{\hat{\theta}}B^{\hat{\phi}}\rho\varpi}{d
u_{\phi}/dr}. \label{mplus}
\end{eqnarray}  In the non-relativistic limit  $u_{\phi}$ can be  identified
as non-relativistic Keplerian angular momentum: $ \Omega_K =
\sqrt{M/r^3}$ and we get
\begin{eqnarray}
\dot{M}_+ &\rightarrow&
\frac{B^{\hat{\theta}}B^{\hat{\phi}}r^2}{r\Omega_K/2} = 2r
\frac{B^{\hat{\theta}}B^{\hat{\phi}}}{\Omega_K},  \end{eqnarray}
which is identical to that used in \cite{lk} and \cite{love}.

Similarly the energy flux of electromagnetic field in
$\theta$-direction is given by
\begin{eqnarray}
\Delta E &=& 2 \int^{r + \delta r}_r \alpha {\cal E}^{\theta}
d\Sigma_{\theta} = - \Omega_F
B^{\hat{\theta}}B^{\hat{\phi}}\rho\varpi \delta r \label{deltae}
\\ &=& \Omega_F \Delta L, \end{eqnarray} where factor 2 is introduced for the
same reason as in the angular momentum flux.
 It is balanced by the change of the
energy of the matter given by
\begin{eqnarray}
\Delta E_m &=& - (-u_{0})\dot{M}_+|^{ r + dr}_{r} \rightarrow
\frac{d u_{0}}{dr}\dot{M}_+ \delta r, \end{eqnarray} which is
possible with $\Omega_F$ given by
\begin{eqnarray}
\Omega_F = - \frac{d u_{0}/dr}{d u_{\phi}/dr},
\label{dudr}\end{eqnarray} where eq.(\ref{mplus}) has been used.

It is not a trivial task to find the solution of the full
relativistic MHD equations satisfying eq.(\ref{M+0}) and
(\ref{mplus}) which leads to
\begin{eqnarray}
 B^{\hat{\theta}} B^{\hat{\phi}} = - 2\pi \sigma_m \rho u^r \frac{d
u_{\phi}}{dr},\label{rconst} \end{eqnarray} together with
eq.(\ref{dudr}). Without solving the full MHD equation, the
magnetic fields and its angular velocity $\Omega_F$ can be
considered to be unknown parameters as well as $u_{i}$.

 As a first trial,
we assume $u_{0}$ and $u_{\phi}$ to be  those of the stable orbit
of a test particle around a Kerr black hole\cite{st} :
\begin{eqnarray}
-u_0 &=& \frac{r^2 -2Mr + a\sqrt{Mr}}{r\sqrt{r^2 -3Mr
+2a\sqrt{Mr}}},\nonumber \\ u_{\phi} &=& \frac{\sqrt{Mr}(r^2
-2a\sqrt{Mr} +a^2)}{r\sqrt{r^2 -3Mr +2a\sqrt{Mr}}}, \label{uphi}
\end{eqnarray} although it is natural to expect non-negligible effect of the
magnetic field on the stable orbit. The straightforward
calculation, eq.(\ref{dudr}), shows a very interesting result that
the angular velocity $\Omega_F$ of the magnetic field is the same
as the Keplerian angular velocity $\Omega_K(\equiv
u^{\phi}/u^{0})$
\begin{eqnarray}
\Omega_F = \Omega_K, \label{fk}\end{eqnarray} where
\begin{eqnarray}
\Omega_K = \sqrt{\frac{M}{r^3}}\frac{1}{1 +
a\sqrt{M/r^3}}.\label{kepler} \end{eqnarray}  In fact one can
derive eq.(\ref{fk}) formally without using the explicit form of
eq.(\ref{uphi}), since it can be  shown  that
\begin{eqnarray}
u^0 \frac{du_0}{dr} +  u^{\phi} \frac{du_{\phi}}{dr} =0
\end{eqnarray} holds in general  for a stable circular orbit on
the equatorial plane. This shows that the energy balance can be
realized by the magnetic field rotating rigidly with Keplerian
angular velocity in this simplified two-dimensional disk model.

It is also possible to show that eq.(4.2) in \cite{blandford} can
be obtained from eq.(\ref{mplus}) in the non-relativistic limit.
Hence for a Newtonian limit or for the outer radius of the thin
disk, the configuration of the electromagneic field suggested by
Blandford\cite{blandford} is consistent  with our result when the
the magnetic field angular velocity $\Omega_F$ is given by
$\Omega_K$.

\section{Power out of a magnetically dominated thin
 accretion disk}

The total power out of the disk, $P_{disk}$, can be calculated by
integrating eq. (\ref{deltae}) from the inner most stable
point($r_{in}$) to  outer most edge of the disk with Poynting
flux($r_{out}$),
\begin{eqnarray}
P_{disk} = - \int_{r_{in}}^{r_{out}}  (-\Omega_F
B^{\hat{\theta}}B^{\hat{\phi}}\rho\varpi) dr, \end{eqnarray} where
$-$ sign is referring the outward direction on the disk.

Using the accretion rate eq.(\ref{mplus}), which is independent of
$r$ for a stationary accretion flow, and the energy balance
condition eq.(\ref{dudr}), we get
\begin{eqnarray}
P_{disk} &=& \int_{r_{in}}^{r_{out}} \Omega_F \dot{M}_+ \frac{d
u_{\phi}}{dr} dr =  \dot{M}_+ \int_{r_{in}}^{r_{out}} \frac{d
(-u_{0})}{dr} dr, \\ &=& [-u_0(r_{out})]\dot{M}_+ -
[-u_0(r_{in})]\dot{M}_+. \label{powerp}\end{eqnarray} The first
term corresponds to the energy rate into the disk at the outer
most edge of the disk and the second term corresponds to the
energy accretion rate at the inner edge of the disk to the central
object.  This is the expected result from the energy conservation.

Eq.(\ref{powerp}) indicates that  the power out of the disk in
this toy model is not dependent strongly on the details of the
electromagnetic field configuration. The power can be calculated
once $\dot{M}_+$ is known at one point $r$ on the disk.  For
example, suppose that the exact solution of the relativistic
equation can be approximated   at a large distance $r_0$ by the
configuration suggested by Blandford\cite{blandford}, then we can
use the relation, for example,
\begin{eqnarray}
B^{\hat{\phi}} = 2 \Omega_F r B^{\hat{\theta}}, \end{eqnarray} to
get the accretion rate expressed in terms of the magnetic field
component perpendicular to the disk,
\begin{eqnarray}
\dot{M}_+ = 4  (B^{\hat{\theta}}(r_0)r_0)^2. \end{eqnarray} Then
the total power for a disk with Poynting flux extended to $r_{out}
= \infty$ can be given by
\begin{eqnarray}
P_{disk} = 4(1 - [-u_0(r_{in})])(B^{\hat{\theta}}(r_0)r_0)^2,
\label{powerp1}\end{eqnarray} where we take $-u_0(\infty) = 1$.

\section{Discussion}

In this work, we discuss a toy model for the magnetically
dominated thin accretion disk. Assuming a two-dimensional disk
dominated by the Poynting flux where the viscous stress and
radiative transfer are ignored, the accretion flow  in
two-dimension is discussed in the background of Kerr geometry. We
have demonstrated that the stationary accretion can be possible
with the co-rotating magnetic field with the same Keplerian
angular velocity as that of matters in the disk in the stable
orbit. Also it is observed that the solution proposed by
Blandford\cite{blandford} is consistent with our result in the
non-relativistic limit( or at a large distance $r$ from the center
with sufficiently small $a/r$ and $a/M$).  It is shown in this toy
model that the total Poynting power out of the disk depends on the
accretion rate and the energy at the inner most  stable orbit but
not on the details of the electromagnetic field configuration.

Related issues to be discussed in the future are the possible
solutions of magnetic field configuration(not only the poloidal
component of the magnetic field discussed by Ghosh\cite{ghosh} but
also the toroidal component) analogous to \cite{blandford} in the
non-relativistic limit and the effect of the magnetic field on the
energy($-u_0$) and angular momentum($u_{\phi}$) of a particle in
the stable orbit which has not been taken into account in this
work.

 We have thus far discussed only one specific aspect, Poynting
flux, of the accretion disk.  For this two-dimensional disk model
to be realistic and viable one, there are many obstacles to
overcome. For example the viscous stress and radiation transfer
should be included, which naturally leads to the questions on the
steady state\cite{love} and  the  thin disk approximation assumed
in this work. Moreover recent
works\cite{gammie}\cite{krolik}\cite{punsly} on the accretion flow
toward a black hole as a central object and on the magnetic
coupling\cite{li} between the disk and the black hole seem to
indicate that the physics is more complex than the simplified
two-dimensional model discussed in this work particularly near the
inner region of the disk.

\vspace*{1cm}

\begin{center}
{\rm\bf Acknowledgements}
\end{center}

The author would like to thank Hongsu Kim and Chul H. Lee for
valuble discussions on the stable orbit.   This work is supported
in part by Hanyang University, Korea, made in the program year of
2000.

\vspace*{2cm}

\begin{center}
{\rm\bf Appendix A : Kerr geometry }
\end{center}

In this Appendix, the useful identities expressed in terms of
explicit orthonormal components defined by ZAMO\cite{TPM} are
listed. Using the
 Boyer-Lindquist coordinates\cite{bl} in the
 natural unit $G=c=1$,  the metric tensor for Kerr Geometry\cite{kerr},
 $g_{ \mu \nu}$, is given by
\begin{eqnarray}
(g_{\mu\nu}) = \pmatrix{-(\alpha^2 -\varpi^2 \beta^2)& 0 & 0
&\varpi^2\beta \cr
 0& \frac{\rho^2}{\Delta} & 0 & 0 \cr
 0 & 0& \rho^2 & 0 \cr
 \varpi^2 \beta & 0 & 0 & \varpi^2 \cr}, \nonumber
\end{eqnarray}
where
\begin{eqnarray}
\alpha &=& \frac{\rho \sqrt{\Delta}}{\Sigma}, ~~
\beta=\frac{g_{0\phi}}{g_{\phi \phi}}, ~~
\tilde{\omega} = \frac{\Sigma}{\rho} \sin \theta, \label{tomega}
\\ \Delta &=& r^2 + a^2 -2Mr, ~~ \rho^2 = r^2 + a^2 \cos^2 \theta,
~~ \Sigma^2 = (r^2 + a^2)^2 - a^2 \Delta \sin^2\theta,
 \end{eqnarray}
ZAMO's four velocity(a time-like unit vector orthogonal to the
t-constant surface: $dx^{\alpha} U_{\alpha} =0$) is given by
\begin{eqnarray}
U_{\mu} = -\alpha(1, \, 0, \, 0, \, 0).  \end{eqnarray}

The electromagnetic field tensor can be expressed by  the electric
and magnetic fields as given by
\begin{eqnarray}
F^{\theta}_{~\phi} &=& \frac{\varpi}{\rho} B^{\hat{r}}, ~~
F^{\theta}_{~ r}   = -\frac{1}{\sqrt{\Delta}} B^{\hat{\phi}}, ~~
F^{\theta}_{~ 0} = \frac{1}{\rho^2}(\alpha \rho E^{\hat{\theta}} +
\beta \rho \varpi B^{\hat{r}}),
\\ F^{0 r} &=& \frac{\sqrt{\Delta}}{\alpha \rho}E^{\hat{r}}, ~~ F^{0
\phi} = \frac{1}{\alpha\varpi} E^{\hat{\phi}}, ~~ F^{\phi r} =
-\frac{\beta \sqrt{\Delta}}{\alpha \rho}E^{\hat{r}} +
\frac{\sqrt{\Delta}}{\varpi \rho} B^{\hat{\theta}}. \end{eqnarray}

\begin{center}
{\rm\bf Appendix B : Surface current on a two-dimensional
accretion Disk}
\end{center}

Consider  a conserved current ${\cal J}^{\mu}$ defined by
\begin{eqnarray}
{\cal J}^{\mu} = J^{\mu} Y(\theta- \pi/2) + j^{\mu},
\end{eqnarray} where the conserved current  $J^{\mu}$ is the bulk
current density satisfying the Maxwell equation,
\begin{eqnarray}
F^{\mu\nu}_{\,\,\,; \nu} = 4\pi J^{\mu}, ~~~ J^{\mu}_{\,\,;\mu}
=0\label{maxwellb} \end{eqnarray} and $Y(\theta-\pi/2)$ satisfies
\begin{eqnarray}
Y(\theta-\pi/2)_{,\mu}  = - \delta_{ \theta \mu}
\delta(\theta-\pi/2) \label{ytheta} \end{eqnarray} From the
conservation of the current,
\begin{eqnarray}
{\cal J}^{\mu}_{\,\,;\mu} =0, \end{eqnarray} we get,
\begin{eqnarray}
 j^{\mu}_{\,\,;\mu}= \frac{1}{4\pi} F^{ \theta \mu}_{\,\,\,;
\mu}\delta(\theta-\pi/2). \end{eqnarray} Then  we can identify
\begin{eqnarray}
j^{\mu} =\frac{1}{4\pi} F^{\theta \mu }\delta(\theta- \pi/2),
\end{eqnarray} since $F^{\theta \theta}$ vanishes identically. It
is an analogous expression as in \cite{damour}.

Now the  charge density  defined by
\begin{eqnarray}
\tilde{\rho}_e = \alpha j^0 = \alpha\frac{F^{ \theta 0}}{4\pi}
\delta(\theta-\pi/2), \end{eqnarray}  can be rewritten as surface
charge density $\sigma_e$:
\begin{eqnarray}
\int \tilde{\rho}_e dV \equiv \int \sigma_e
\frac{\rho}{\sqrt{\Delta}} \varpi d r d \phi. \end{eqnarray} Then
we get the surface charge density in terms of  the electric field
given in Appendix A:
\begin{eqnarray}
\sigma_e = -\frac{E^{\hat{\theta}}}{4\pi},\label{eE1}
\end{eqnarray} It is Gauss' law on the disk plane:
\begin{eqnarray}
E^{\hat{\theta}} = - 4\pi  \sigma_e. \label{gauss} \end{eqnarray}
Similarly using the current density defined by
\begin{eqnarray}
\tilde{j}^{i} = j^i -j^0(0,\,0, \, -\beta), ~~~ i = r, ~ \theta, ~
\phi
 \end{eqnarray}
  we get
 \begin{eqnarray}
 \tilde{j}^{r} = -\frac{1}{4\pi}\frac{B^{\hat{\phi}}\sqrt{\Delta}}{\rho^2}
 \delta(\theta-\pi/2), ~~~
 \tilde{j}^{\phi} =
 \frac{1}{4\pi}\frac{B^{\hat{r}}}{\rho \varpi} \delta(\theta-\pi/2)
 \label{tildej},\end{eqnarray}
where $j^{\theta} =0$ by construction. The radial surface current
density $K^{\hat{r}}$ and the surface current density in $\phi-$
direction $K^{\hat{\phi}}$ on the disk can be defined by
\begin{eqnarray}
 \int \alpha \tilde{j}^r d\Sigma_{r}  &=& -\frac{1}{4\pi}
  \int \alpha B^{\hat{\phi}} \varpi d \phi
 \equiv \int \alpha
 K^{\hat{r}} \varpi d \phi, \\
\int \alpha \tilde{j}^{\phi} d\Sigma_{\phi} &=&
\frac{1}{4\pi}\int \alpha B^{\hat{r}} \frac{\rho}{\sqrt{\Delta}} d
r \equiv \int \alpha
 K^{\hat{\phi}} \frac{\rho}{\sqrt{\Delta}} d r,
 \end{eqnarray}
to get
\begin{eqnarray}
K^{\hat{r}} = -\frac{1}{4\pi} B^{\hat{\phi}}, ~~~~~ K^{\hat{\phi}}
=\frac{1}{4 \pi} B^{\hat{r}},\label{kbd1} \end{eqnarray} where
$d\Sigma_i$ is the corresponding surface element.

This result can be summarized by   Ampere's law on the disk
surface:
\begin{eqnarray}
\vec{B} = - 4\pi \vec{K} \times \hat{\theta}.\label{kbad}
 \end{eqnarray}

\end{document}